\documentclass{article}
\usepackage{amsmath}
\usepackage{amssymb}
\usepackage{graphicx}
\begin{document}
\begin{center}
\LARGE
\textbf{A generic approach to the quantum mechanical transition 
probability}
\vspace{1 cm}
\normalsize

Gerd Niestegge
\vspace{0,1 cm}
\footnotesize

gerd.niestegge@web.de
\vspace{0,5 cm}
\end{center}
\normalsize
\begin{abstract}
\noindent
In quantum theory, the modulus-square of the inner product 
of two normalized Hilbert space elements is to be interpreted 
as the transition probability between the pure states represented 
by these elements. A probabilistically motivated and 
more general definition of this transition probability 
was introduced in a preceding paper and is 
extended here to a general type of quantum logics: 
the orthomodular partially ordered sets. A very general 
version of the quantum no-cloning theorem, creating promising 
new opportunities for quantum cryptography, is presented and 
an interesting relationship between the transition probability 
and Jordan algebras is highlighted.
\vspace{0,5 cm}

\end{abstract}
\vspace{0,5 cm}

\section{Introduction}

Max Born’s statistical interpretation \cite{born1926} made probability 
play a major role in quantum theory. He postulated that
the modulus-square of the inner product
of two normalized Hilbert space elements
should be interpreted as a transition probability
between the pure states represented by the two
Hilbert space elements. The mathematical formalism
does not provide any reason for this interpretation,
but the experimental evidence forces us to accept it.

Since then various approaches have emerged to find 
a better motivated axiomatic access to Born's postulate, using
convex sets \cite{AS02, mielnik1969theory, mielnik1974generalized}, 
transition probability spaces \cite{Landsman1997, pulmannova1989ql_and_trans_prob},
continuous geometries \cite{vNgeotp}
or quantum logics \cite{guz1980non, maczynski1981}.
Among these, only the convex set approach does not postulate
the existence of the transition probability
by an extra axiom.
Here we use the quantum logics,
which were originally pioneered by Birkhoff and von Neumann \cite{birkhoff-vN36}.

Earlier quantum logical approaches by the author were based on projective 
quantum measurement (L\"uders~-~von Neumann quantum measurement process) 
or on an extension of the classical
conditional probability \cite{Nie1998HPA01, niestegge2001non}. 
A different approach 
was undertaken in a previous paper~\cite{nie2020alg_origin}.
Its intention was to
point to the algebraic origin of the quantum probabilities
and their inherent difference 
from the classical probabilities.
Therefore, only the Hilbert space quantum logics 
were considered, and the transition probability 
was defined in a new way. The new definition
includes physically meaningful and
experimentally verifiable novel cases
that are covered neither by the usual quantum mechanical
transition probability nor by the approaches of other authors
\cite{AS02,guz1980non, Landsman1997, maczynski1981, mielnik1969theory, mielnik1974generalized, 
pulmannova1989ql_and_trans_prob, vNgeotp}.

The extension of this definition
to general quantum logics (\emph{orthomodular partially ordered sets} 
\cite{gudder1979stochastic, ptak1991orthomodular, varadarajan1968and1970})
is quite straight forward and is studied in the present paper.
This transition probability does neither require any extra axiom
nor the extended conditional probability,
but is a characteristic of the algebraic structure of the quantum logic; 
in some cases it exists and in others it does not.
A very general version of the quantum \emph{no-cloning theorem} 
\cite{dieks1982communication, wootters1982single}
is presented and proven in this setting, which creates promising
new opportunities for the quantum key distribution protocols \cite{BB84, E91}
- even in the common quantum mechanical Hilbert space setting.
An interesting class of quantum logics are
the projection lattices of the
\emph{JBW-algebras} (the Jordan
analogue of the von Neumann algebras or W*-algebras 
\cite{AS02, hanche1984jordan, sakai1971}); 
it is shown how the transition probability 
is linked to the algebraic structure in this case
and that the Jordan algebras provide the appropriate
framework for a structural analysis of the 
transition probability.

This paper requires some knowledge of two 
mathematical subjects: 
the orthomodular partially ordered sets and 
the JBW-algebras. Some basics 
needed in the paper
are briefly sketched in sections 2 and 6, respectively.
Beyond that, it is referred to the monographs 
\cite{ptak1991orthomodular, varadarajan1968and1970} 
for the first subject and to the monographs 
\cite{AS02, hanche1984jordan} for the second one.

The paper is organized as follows. 
In section 2, some basics of the 
orthomodular partially ordered sets 
(the general mathematical model for the quantum logic),
the states on them and their morphisms
are recapitulated for later use.
In section 3, the transition probability is 
introduced and its basic properties 
are presented. The relation between 
this transition probability and 
the quantum logical
notion of \emph{compatibility}
is studied in section 4. Its main result 
is a certain product rule for the 
transition probability and becomes the
major tool for the derivation of 
the no-cloning theorem in section~5.
In section 6, some basics of the 
JBW-algebras and their 
associated quantum logics
are recapitulated for later use.
How the transition probability in these
quantum logics is linked to the algebraic structure
is studied in section~7. 
The relation between the transition probability defined here 
and the usual transition probability in the common quantum mechanical
Hilbert space formalism
can then easily be disclosed. 
Moreover, it becomes apparent that
a certain property known for the von Neumann algebras
as \emph{isoclinicity} \cite{christensen1982measures, hamhalter2015dye, maeda1989probability} 
or \emph{equiangularity} \cite{Freedman_2019}
is very closely related to the 
transition probability. This property is then
studied in the Jordan algebraic framework in section 8, 
the results of which are used in section~9 for the structural 
analysis of the transition probability in the 
Jordan algebraic quantum logics. Theorem 9.3 and, prior to this,
the product rule (section 4), the no-cloning theorem (section~5) 
and the link between the transition probability and 
the Jordan algebraic structure (section~7) 
represent the main results of the paper.

\section{Quantum logics and states}

The Boolean algebra is a mathematical structure playing an important role in many scientific and technical fields such as formal logic, classical probability theory, circuitry, computer science. Only quantum theory challenges the general applicability of this structure, since the dichotomic observables (those with spectrum $\left\{0,1\right\}$) 
do not form a Boolean algebra, but an orthomodular lattice where the distributivity law fails \cite{beltrametti1984logic, birkhoff-vN36, kalmbachorthomodular, varadarajan1968and1970}.
The dichotomic observables are identical with the self-adjoint projections
and can also be identified with the closed linear subspaces of the Hilbert space.
The new mathematical structure which they form is generally called a \emph{quantum logic}.

The lattice structure used to play an important role in 
the early quantum logical approaches \cite{birkhoff-vN36, piron1964axiomatique,
varadarajan1968and1970}. 
However, there is no physical motivation 
for the existence of the lattice operations
for elements of the quantum logic
that are not compatible, and later a quantum logic 
was often assumed to be an orthomodular partially ordered set only
\cite{gudder1979stochastic, ptak1991orthomodular, ptak1983measures}.
This more general structure will also be sufficient here.

A \textit{quantum logic} shall be an \emph{orthomodular partially ordered set} $L$ 
with order relation $\leq$, 
smallest element $0$, largest element $\mathbb{I}$ and an ortho\-complement\-ation~$'$.
This means that the following conditions are satisfied by the $p,q \in L$:
\begin{enumerate}
\item[(a)] $ q \leq p$ \textit{implies} $p' \leq q'$.
\item[(b)] $(p')' = p$.
\item[(c)] $p \leq q'$ \textit{implies that} $p \vee q$, \textit{the supremum of} $p$ \textit{and} $q$, \textit{exists}.
\item[(d)] Orthomodular law: $q \leq p$ \textit{implies} $p = q \vee (p \wedge q')$.
\end{enumerate}
Here, $p \wedge q$ denotes the infimum of $p$ and $q$, 
which exists iff $p' \vee q'$ exists. 
Note that $p \vee p' = \mathbb{I}$ holds for $p \in L$; this follows from (d).

In a lattice, 
$p \wedge q$ and $p \vee q$ would exist for any elements $p$ and $q$.
An element $e \in L$ with $e \neq 0$ is called \textit{minimal} 
if there is no $q \in L$ with $q \leq e$ and $0 \neq q \neq e$.
The minimal elements are also called \textit{atoms} in the common literature.
Two elements $p$ and $q$ in $L$ are \textit{orthogonal}, 
if $p \leq q'$ or, equivalently, $q \leq p'$;
in this case, $p \vee q$ exists and shall be noted by $p + q$ in the following.
Moreover, if $p$ and $q$ are orthogonal,
$p \wedge q$ exists and $p \wedge q = 0$. In a Boolean algebra,
the identity $p \wedge q = 0$ is the same as the orthogonality 
of $p$ and $q$. However, this does not hold any more in a quantum logic.

The interpretation of this mathematical terminology is as
follows: orthogonal elements represent mutually exclusive potential 
measurement outcomes, $p'$ represents the negation of
$p$, $p + q := p \vee q $ is the disjunction
(logical \emph{or}-operation) of $p$ and $q$
which generally exists only if these two elements are orthogonal,
and $\leq$ is the logical implication relation ($q \leq p$ with $q,p \in L$
means that the measurement outcome $q$ implies the outcome $p$).

A \textit{state} $\mu$ shall allocate probabilities to the 
elements of the quantum logic. Therefore it becomes a map
from $L$ to the unit interval $\left[0,1\right] \subseteq \mathbb{R}$
with $\mu(\mathbb{I})=1$ and $\mu(p + q) = \mu(p) + \mu(q)$ for any two
orthogonal elements $p$ and $q$ in $L$.
A set $S$ of states on $L$ is called 
\textit{strong} if, for any $p,q \in L$,
$$ \left\{ \mu \in S \:|\: \mu(q) = 1 \right\} \subseteq 
\left\{ \mu \in S \:|\: \mu(p) = 1 \right\} \ \  \Rightarrow \ \  q \leq p .$$
Note that a strong set $S$ contains 
a state $\mu \in S$ with $\mu(q) = 1$ 
for each $q \in L$ with $q \neq 0$; if 
$\left\{ \mu \in S \:|\: \mu(q) = 1 \right\} = \emptyset$, 
we would get $q \leq p$ for all $p \in L $ and thus $q = 0$.

A \textit{morphism} from a quantum logic $L$ 
to another quantum logic $K$ 
is a map $\pi : L \rightarrow K$ satisfying 
the following two conditions:
\begin{enumerate}
\item[(a)] $\pi(\mathbb{I}) = \mathbb{I}$
\item[(b)] If $p \in L$ and $q \in L$ are orthogonal, then 
$\pi(p)$ and $\pi(q)$ are orthogonal in $K$ 
and $\pi(p + q) = \pi(p) +\pi(q)$.
\end{enumerate}
With a state $\mu$ on $K$, a state 
$\mu \pi : L \ni p \rightarrow \mu(\pi p)$ 
can then be defined on $L$.
However, if $S$ is a certain set of states on $L$,
$S'$ a certain set of states on $K$ and $\mu \in S'$, 
the state $\mu \pi$ need not lie in $S$. Therefore,
an $S$-$S'$-\textit{morphism} becomes a morphism 
satisfying the following additional condition:
\begin{enumerate}
\item[(c)] If $\mu \in S'$, then $\mu \pi \in S$.
\end{enumerate}
In most interesting cases, 
$S$ will be the set of \emph{all} states on $L$
and the additional condition (c) will be needless, but a few cases
will require the use of a smaller set of states, as we shall see later.

\section{Transition probability in quantum logics}

In a preceding paper \cite{nie2020alg_origin},
the \emph{transition probability} was defined in a new way 
for the Hilbert space quantum logics, 
but its following extension 
to the general situation is quite straightforward.
\vspace{0,3 cm}

\textbf{Definition 3.1} 
\itshape 
Let $L$ be a quantum logic and $S$ a strong set of states on $L$.
If a pair $p,q \in L$ with $p \neq 0$ and some $s \in [0,1]$
satisfy the identity 
\begin{center}
$\mu(q) = s$ for all $\mu \in S$ with $\mu(p)=1$,
\end{center}
then $s$ is called the \emph{transition probability from} $p$ \emph{to} $q$ 
and is denoted by $\mathbb{P}(q|p)$.
\normalfont
\vspace{0,3 cm}

The identity
$\mathbb{P}(q|p) = s$ then becomes equivalent to the set inclusion
\begin{equation*}
\left\{\mu \in S \:|\: \mu(p)=1\right\} \subseteq \left\{\mu \in S \:|\: \mu(q)=s\right\}
\end{equation*}
and means that, whenever the probability of $p$ is $1$, 
the probability of $q$ is fixed and its numerical value is $s$;
particularly in the situation after a quantum measurement
that has provided the outcome $p$, the probability of $q$ 
becomes $s$, independently of any initial state 
before the measurement.

Two elements $p$ and $q$ in $L$ 
are orthogonal iff $\mathbb{P}(q | p) = 0$,
and $p \leq q$ holds iff $\mathbb{P}(q | p) = 1.$
The second part here holds since $S$ is a strong set of states,
and the first part follows by considering $q'$.

The following lemma provides a collection of 
some further basic properties of the transition probability.
\vspace{0,3 cm}

\textbf{Lemma 3.2} 
\itshape 
Let $L$ be a quantum logic and $S$ a strong set of states on~$L$.
\begin{enumerate}
\item[\normalfont(i)]
If $\mathbb{P}(q | p)$ exists and $0 \neq p_o \leq p$ with $p_o,p,q \in L$, then
$\mathbb{P}(q | p_o)$ exists and $ \mathbb{P}(q | p_o) = \mathbb{P}(q | p)$.
\item[\normalfont(ii)]
If $\mathbb{P}(q | p)$ exists for $p,q \in L$ and $ \mathbb{P}(q | p) \neq 1 $,
then $p \wedge q = 0$.
\item[\normalfont(iii)]
If $0 \neq q < p$ with $p,q \in L$, then $\mathbb{P}(q | p)$ does not exist.
\item[\normalfont(iv)]
If $\mathbb{P}(q | p)$ exists for all $q \in L$ 
with $0 \neq p \in L$ (in this case, $p$ defines
a state $L \ni q \rightarrow \mathbb{P}(q | p)$), 
then $p$ is an atom (a minimal element in $L$).
\item[\normalfont(v)]
Let $K$ be a second quantum logic, 
let $S'$ be a strong set of states on $K$, and 
let $\pi: L \rightarrow K$ be a $S$-$S'$-morphism. If 
$\mathbb{P}(q | p)$ exists for $0 \neq p$ with $p,q \in L$, 
then $\mathbb{P}(\pi q | \pi p)$ 
exists and $\mathbb{P}(\pi q | \pi p) = \mathbb{P}(q | p)$.
\end{enumerate}
\normalfont
Proof. 
(i) Suppose $\mathbb{P}(q | p)$ exists and 
$0 \neq p_o \leq p$ with $p_o,p,q \in L$. Then
$$\left\{ \mu \in S|\mu(p_o) = 1 \right\} \subseteq 
\left\{ \mu \in S|\mu(p) = 1 \right\} \subseteq 
\left\{ \mu \in S|\mu(q) = \mathbb{P}(q | p) \right\}.$$
This means $ \mathbb{P}(q | p_o) = \mathbb{P}(q | p)$.

(ii) Suppose $p_o \leq p$, $p_o \leq q$ and $p_o \neq 0$ 
with $p_o,p,q \in L$.
If $\mathbb{P}(q | p)$ exists, we get by (i): 
$\mathbb{P}(q | p) = \mathbb{P}(q | p_o) = 1$.

(iii) Suppose $0 \neq q < p$ with $p,q \in L$. 
The orthomodularity implies that 
$p \wedge q' \neq 0$ and, 
since $S$ is strong, there are states
$\mu_1, \mu_2 \in S$ with $\mu_1(q) = 1$ and $\mu_2(p \wedge q') = 1$. 
Then $\mu_1(p) = \mu_2(p) = 1$, but $\mu_2(q) = 0 \neq 1 = \mu_1(q)$.

(iv) If there were an element $q_o \in L$ with
$0 < q_o < p \in L$, then $\mathbb{P}(q_o | p)$ does not exist by (iii).

(v) Suppose that $\mathbb{P}(q | p)$ exists 
for $p,q \in L$ with $p \neq 0$ and that
$\mu \pi p = 1$ holds for $\mu \in S'$. Then $\mu \pi \in S$
and thus $\mu \pi q = \mathbb{P}(q | p)$. Therefore
$\mathbb{P}(\pi q | \pi p)$ 
exists and $\mathbb{P}(\pi q | \pi p) = \mathbb{P}(q | p)$.
\hfill $\square$
\vspace{0,3 cm}

When the transition probability $\mathbb{P}(q | p)$ exists
for two elements $p \neq 0$ and $q$ in the quantum logic $L$,
part (i) of Lemma 3.2
means that, after a quantum measurement 
with the outcome $p$, a further measurement of any $p_o \in L$ with $0 \neq p_o \leq p$ 
cannot alter the probability of $q$.

We shall see later in Corollary 7.2 that,
in many important cases, the reverse implication of (iv) 
is also true, and a unique state can then be allocated to each atom
such that the atom carries the probability $1$. 

Part (v) of Lemma 3.2 shows that the transition probability is
invariant under morphisms.

The relation between the transition probability defined here
and the quantum mechanical transition probability
$ \left|\left\langle  \xi | \psi \right\rangle\right|^{2} $
for two normalized Hilbert space elements 
(or wave functions or pure states)
$\xi$ and $\psi$
is not obvious and has been elaborated 
in Ref. \cite{nie2020alg_origin}.
We shall come back to this in section 7.

\section{Compatibility}

Two elements $p$ and $q$ in a quantum logic $L$ 
are said to be \textit{compatible},
if there exist three pairwise orthogonal elements 
$a_1, a_2, a_3 \in L$ such that $p = a_1 + a_2$ and $q = a_2 + a_3$;
it is well-known 
that the elements $a_1, a_2, a_3$ are unique if they exist,
namely $a_1 = p \wedge q'$, $a_2 = p \wedge q$ and $a_3 = q \wedge p'$
\cite{brabec1979compatibility, ptak1991orthomodular}.
Here, $\wedge$ can be interpreted as the logical \emph{and}-operation.
However, this interpretation is not reasonable
for the infimum of two non-compatible elements when it exists.
Note that two elements $p$ and $q$ are compatible in the cases when
they are orthogonal or when $p \leq q$ or $q \leq p$.

It is well-known and easy to see that two elements $p$ and $q$ 
in the usual Hilbert space quantum logic of
common quantum mechanics are compatible iff they commute;
in this case $p \wedge q = pq = qp$.
\vspace{0,3 cm}

\textbf{Lemma 4.1} 
\itshape
Let $L$ be a quantum logic and $S$ 
a strong set of states on $L$.
If $0 \neq p \in L$ and $q \in L$ are compatible
and if $\mathbb{P}(q | p)$ exists,
then either $p$ and $q$ are orthogonal and $\mathbb{P}(q | p) = 0$,
or $p \leq q$ and $\mathbb{P}(q | p) = 1$.
\normalfont
\vspace{0,3 cm}

\textbf{Proof.} Suppose that $\mathbb{P}(q | p)$ exists
for $p,q \in L$, $p\neq0$, and let $a_1, a_2, a_3$ 
be three pairwise orthogonal elements in $L$
with $p = a_1 + a_2$ and $q = a_2 + a_3$.
If $a_2 = 0$, $p$ and $q$ are orthogonal and $\mathbb{P}(q | p) = 0$.
If $a_2 \neq 0$, there is a state $\mu \in S $ with $\mu(a_2) = 1$
(since $S$ is a strong set of states);
then $\mu(p) = 1$ and $\mathbb{P}(q | p) = \mu(q) = 1$.
\hfill $\square$
\vspace{0,3 cm}

Lemma 4.1 means that a non-trivial transition probability
requires incompatibility and does not exist in the classical 
logics which are the Boolean algebras and where all elements 
are compatible with each other.

Two elements $p$ and $q$ in a quantum logic $L$ are compatible 
iff there exists a Boolean subalgebra of $L$ 
containing $p$ and $q$.
However, if three or more elements are pairwise compatible,
they need not lie in a joint Boolean subalgebra of $L$ in general.
They do so if $L$ is a lattice \cite{brabec1979compatibility, varadarajan1968and1970}.
In the general case, we have only the following lemma
stating that 
three elements lie in a joint Boolean subalgebra of $L$, 
if they are pairwise compatible and two among them
are orthogonal.
\vspace{0,3 cm}

\textbf{Lemma 4.2} 
\itshape
If $p$, $q_1$ and $q_2$ are three elements 
in a quantum logic $L$ such that
$q_1$ and $q_2$ are orthogonal,
$p$ and $q_1$ are compatible and
$p$ and $q_2$ are compatible,
then $p$ and $q_1 + q_2$ are compatible and we have
$$p \wedge (q_1 + q_2) = p \wedge q_1 + p \wedge q_2 .$$
\normalfont

\textbf{Proof.} 
Suppose $p, q_1, q_2 \in L$ such that $q_1$ and $q_2$ are orthogonal,
$p$ and $q_1$ are compatible and
$p$ and $q_2$ are compatible.
The five elements
$p \wedge q_1 $, $p \wedge q_2 $, $p' \wedge q_1 $, $p' \wedge q_2 $ and 
$p \wedge (q_1 + q_2)'$ are then pairwise orthogonal 
and therefore they lie in a Boolean subalgebra of $L$ 
\cite{brabec1979compatibility, brabec-ptak1982compatibility,
dorninger2014, ptak1991orthomodular}.
It includes 
$ p = p \wedge q_1 + p \wedge q_{1}' = p \wedge q_2 + p \wedge q_{2}'$,
$q_1 = p \wedge q_1 + p' \wedge q_1$ and $q_2 = p \wedge q_2 + p' \wedge q_2$.
This implies the the compatibility of $p$ and $q_1 + q_2$ 
and the above identity. \hfill $\square$
\newpage

Now we shall see that the compatibility implies 
a certain product rule for the transition probability.
\vspace{0,3 cm}

\textbf{Proposition 4.3}
\itshape
Let $K_1,K_2,L$ be three quantum logics and let $S_1$, $S_2$ and $S$ 
be strong sets of states for them. Moreover, suppose that 
$K_1 \ni p_1 \rightarrow \bar{p_1} \in L$ is an $S_1$-$S$-morphism and that
$K_2 \ni p_2 \rightarrow \tilde{p_2} \in L$ is an $S_2$-$S$-morphism
such that $\bar{p_1}$ and $\tilde{p_2}$ are a compatible pair in $L$ 
for each $p_1 \in K_1$ and each $p_2 \in K_2$.

If $\mathbb{P}(q_1 | p_1)$ and $ \mathbb{P}(q_2 | p_2) $ exist 
for some $q_1,p_1 \in K_1$, $q_2,p_2 \in K_2$ 
and if $\bar{p_1} \wedge \tilde{p_2} \neq 0$, 
then $\mathbb{P}(\bar{q_1} \wedge \tilde{q_2} | \bar{p_1} \wedge \tilde{p_2} )$ 
exists and
$$\mathbb{P}(\bar{q_1} \wedge \tilde{q_2} | \bar{p_1} \wedge \tilde{p_2} ) 
= \mathbb{P}(q_1 | p_1) \mathbb{P}(q_2 | p_2) .$$
\normalfont

\textbf{Proof}. Suppose that
$\mathbb{P}(q_1 | p_1)$ and $ \mathbb{P}(q_2 | p_2) $ exist 
with $q_1,p_1 \in K_1$, $q_2,p_2 \in K_2$ 
and $\bar{p_1} \wedge \tilde{p_2} \neq 0$.
Let $\mu$ be a state on $L$ with $\mu(\bar{p_1} \wedge \tilde{p_2}) = 1$.
We shall show that 
$\mu(\bar{q_1} \wedge \tilde{q_2}) = \mathbb{P}(q_1 | p_1) \mathbb{P}(q_2 | p_2)$.

Note that $\mu(\bar{p_1}) = 1 = \mu(\tilde{p_2})$ and
define states $\mu_0$ and $\mu_1$ on $K_1$ by 
$\mu_0(q) := \mu(\bar{q})$ and
$\mu_1(q) := \mu(\bar{q} \wedge \tilde{p_2})$ for $q \in K_1$; 
Lemma 4.2 ensures that $\mu_1$ is a state.
Then $ \mu_0(p_1) = 1 = \mu_1(p_1)$ and thus 
$\mu(\bar{q_1}) = \mu_0(q_1) = \mathbb{P}(q_1 | p_1) = \mu_1(q_1)$.

If $ \mathbb{P}(q_1 | p_1) \neq 0$, define a state $\mu_2$ on $K_2$ by 
$$ \mu_2(q) := \frac{1}{\mathbb{P}(q_1 | p_1)} \mu(\bar{q_1} \wedge \tilde{q})$$
for $q \in K_2$; again Lemma 4.2 ensures that $\mu_2$ is a state. Then
$$ \mu_2(p_2) = \frac{1}{\mathbb{P}(q_1 | p_1)} \mu(\bar{q_1} \wedge \tilde{p_2})
= \frac{1}{\mathbb{P}(q_1 | p_1)} \mu_1(q_1) = 1$$
and thus $\mu_2(q_2) = \mathbb{P}(q_2 | p_2)$.
Therefore 
$$ \mathbb{P}(q_2 | p_2) = \frac{1}{\mathbb{P}(q_1 | p_1)} \mu(\bar{q_1} \wedge \tilde{q_2}),$$
which is the desired result.

In the case $ \mathbb{P}(q_1 | p_1) = 0$, $q_1$ and $p_1$ are orthogonal. 
Therefore $\bar{q_1}$ and $\bar{p_1}$ are orthogonal, and then 
$\bar{q_1} \wedge \tilde{q_2}$ and $ \bar{p_1} \wedge \tilde{p_2}$
are orthogonal. Thus 
$\mathbb{P}(\bar{q_1} \wedge \tilde{q_2} | \bar{p_1} \wedge \tilde{p_2} ) 
= 0 = \mathbb{P}(q_1 | p_1) \mathbb{P}(q_2 | p_2) $.
\hfill $\square$

\section{The no-cloning theorem}

Using Lemma 3.2 (v) and Proposition 4.3, we shall now
prove a very general version of the well-known 
quantum no-cloning theorem. This version does neither
require any type of state nor any tensor product.
Instead of the tensor product, just compatibility is sufficient.
Instead of states, elements of a quantum logic are considered; 
they represent properties of a quantum system and potential
measurement outcomes.

The following situation is assumed: A measurement on a system
was performed. However, it is unknown which observable was tested.
Available is only the information that the measurement outcome was 
one among the system properties $p_1$,...,$p_n$,
but it is not known which one. 
It is assumed that the transition probability exists 
for each pair chosen from $p_1$,...,$p_n$.
The following theorem then states that
the unknown property cannot be cloned 
if the properties $p_1$,...,$p_n$ are not pairwise orthogonal.

In the usual quantum mechanical setting, the cloning is 
performed by a unitary transformation on a Hilbert space tensor product
or by the corresponding inner automorphism on the tensor product 
of the operator algebras; in this paper, 
it shall be performed by a morphism from the quantum logic $L$
to itself, where $L$ represents a large system
containing the subsystem where the copy is taken from
and the subsystem where the copy is to be transferred to.
\vspace{0,3 cm}

\textbf{Theorem 5.1}
\itshape
Let $K$ and $L$ be quantum logics and let $S_K$ and $S_L$ 
be strong sets of states for them. Moreover, 
suppose that there are two $S_K$-$S_L$-morphisms from $K$ to $L$:
$K \ni p \rightarrow \bar{p} \in L$ and 
$K \ni p \rightarrow \tilde{p} \in L$
such that $\bar{p}$ and $\tilde{q}$ are a compatible pair in $L$ 
for each $p \in K$ and each $q \in K$ and 
$ \bar{p} \wedge \tilde{q} \neq 0$ for 
for each $p \in K$ and each $q \in K$ with
$p \neq 0 \neq q$.\footnote{
The last assumption 
($ \bar{p} \wedge \tilde{q} \neq 0$ for $p,q \in K$ with $p \neq 0 \neq q$)
means that the two copies of $K$ are
\emph{logically independent} in $L$.
Logical independence is usually defined for 
von Neumann subalgebras \cite{hamhalter1997statistical, redei1995logical} 
and becomes a necessary and sufficient condition 
for the \emph{C*-independence} 
of two commuting von Neumann subalgebras \cite{redei1995logical}; 
C*-independence was introduced by Haag and Kastler in the framework 
of algebraic quantum field theory \cite{haag1964algebraic}.
}

A cloning transformation for $ 0 \neq p_1,...,p_n \in K$
is an $S_L$-$S_L$-morphism $ T : L \rightarrow L $ with 
$$T \left(\bar{p_k} \wedge \tilde{q_o} \right) = \bar{p_k} \wedge \tilde{p_k} $$
for $k = 1,...,n$, where $ 0 \neq q_o \in K$ represents the known initial property
of the second system which shall be replaced 
by the copy of the unknown property 
(one among $p_1,...,p_n$) of the first system.

Suppose that the transition probabilities
$\mathbb{P}\left(p_j|p_k\right)$ exist on $K$ for $j,k = 1,...,n$.
If a cloning transformation $T$ exists for these  $p_1,...,p_n \in K$, 
any two elements chosen from $p_1,...,p_n$ 
must be either orthogonal or identical.
\normalfont
\vspace{0,3 cm}

\textbf{Proof.} Note that both Lemma 3.2 (v) and Proposition 4.3 
are used repeatedly in the following equation; 
Lemma 3.2 (v) is applied with the morphisms $p \rightarrow \bar{p}$, 
$p \rightarrow \tilde{p}$ and the cloning transformation $T$.
\begin{align*}
\left(\mathbb{P}\left(p_j|p_k\right)\right)^{2} 
& = \mathbb{P}\left(\bar{p_j}|\bar{p_k}\right) \mathbb{P}\left(\tilde{p_j}|\tilde{p_k}\right)
  = \mathbb{P}\left(\bar{p_j} \wedge \tilde{p_j} | \bar{p_k} \wedge \tilde{p_k} \right) \\
& = \mathbb{P}\left(T \left( \bar{p_j} \wedge \tilde{q_o} \right) | T \left( \bar{p_k} \wedge \tilde{q_o} \right) \right)
  = \mathbb{P}\left( \bar{p_j} \wedge \tilde{q_o} | \bar{p_k} \wedge \tilde{q_o} \right) \\
& = \mathbb{P}\left(\bar{p_j}|\bar{p_k}\right) \mathbb{P}\left( \tilde{q_o} | \tilde{q_o} \right) 
  = \mathbb{P}\left(\bar{p_j}|\bar{p_k}\right) \\
& = \mathbb{P}\left(p_j|p_k\right)
\end{align*}
and thus $ = \mathbb{P}\left(p_j|p_k\right) \in \left\{0,1\right\}$
for $j,k = 1,...,n$.
If $ = \mathbb{P}\left(p_j|p_k\right) = 1$, then $p_k \leq p_j$ and, 
since $\mathbb{P}\left(p_k|p_j\right)$ exists, we get $p_j = p_k$ by Lemma 3.2 (iii).
If $ = \mathbb{P}\left(p_j|p_k\right) = 0$, $p_j$ and $p_k$ are orthogonal.
\hfill $\square$
\newpage

The rather general and abstract version of the 
no-cloning theorem presented here helps to
identify its deeper origin which is hidden
in the common quantum mechanical Hilbert space formalism 
like the needle in the haystack. We see that only
the existence of the transition probabilities
and two properties of them are needed in the proof:
their invariance under morphisms (Lemma 3.2 (v)) 
and the product rule (Proposition 4.3).
The requirement that the transition probabilities exist 
does not occur in the original quantum no-cloning theorem
\cite{dieks1982communication, wootters1982single},
but it is automatically
fulfilled since only pure states or atoms are considered
and since the transition probabilities always exist for the pure states or atoms
in the usual Hilbert space setting of quantum mechanics
(see Ref.~\cite{nie2020alg_origin} or Corollary 7.2 and the subsequent remarks).

Theorem 5.1 is substantially more general 
than the original no-cloning theorem.
Even in the usual Hilbert space setting, Theorem 5.1
includes physically meaningful interesting new
cases where $ p_1,...,p_n $ are not atomic
and states cannot be allocated. 
This becomes possible by considering the cloning of 
system properties instead of states and by using 
the transition probability $\mathbb{P}(\ |\ )$ 
defined in section 3.

The original no-cloning theorem 
has been extended into different other directions:
to mixed states \cite{barnum1996noncommuting},
to C*-algebras \cite{clifton2003characterizing},
to finite-dimensional generic probabilistic models 
\cite{barnum2006cloning, barnum2007generalized}, and
to universal cloning \cite{:/content/aip/journal/jmp/50/10/10.1063/1.3245811}. 
Possible is only the approximate or imperfect cloning 
\cite{PhysRevA.57.2368, buvzek1996quantum, Kitajima2015}. 
However, none of these extensions includes the above result Theorem 5.1.

The original no-cloning theorem
is essential for the quantum key distribution protocols
\cite{BB84, E91}. How these protocols
can be extended to a much more general setting,
using the above no-cloning Theorem 5.1,
is shown in Ref. \cite{nie2017QKD}. 
The existence of the extended conditional 
probabilities is assumed in Ref. \cite{nie2017QKD}, 
but is not relevant for the key distribution protocols.
It is needed there to derive the transition probabilities,
which have been derived here in a different way 
that does not require the conditional 
probabilities.
Particularly the non-atomic cases of the no-cloning Theorem 5.1
go beyond the usually considered situation and 
create promising new opportunities for the quantum
key distribution protocols - even in common Hilbert 
space quantum mechanics.

\section{The quantum logic of a Jordan algebra}

The \emph{formally real Jordan algebras} were introduced and classified
by Jordan, von Neumann and Wigner \cite{von1933algebraic}. 
Much later, this theory was extended to include infinite dimensional algebras;
these are the so-called \emph{JB-algebras} 
and \emph{JBW-algebras} \cite{AS02, hanche1984jordan},
which represent the Jordan analogue of the C*-algbras 
and the W*-algebras (von Neuman algebras \cite{sakai1971}).
In this section, some basics of the theory of the JB-/JBW-algebras 
shall be recapitulated for later use.

A \textit{real Jordan algebra} is a $\mathbb{R}$-linear space $A$ equipped 
with an abelian (but not associative) product $\circ$ satisfying
$$x^{2} \circ (x \circ y) = x \circ (x^{2} \circ y)$$
for any $x,y \in A$. For any three elements $x,y,z \in A$,
their \textit{triple product} is defined as follows:
$$\left\{x,y,z\right\} = x \circ (y \circ z) - y \circ (z \circ x) + z \circ (x \circ y).$$

A \textit{JB-algebra} is a real Jordan algebra $A$ 
that is a Banach space with a norm satisfying
$ \left\|x \circ y\right\| \leq \left\|x\right\| \left\|y\right\| $, 
$\left\|x^{2}\right\| = \left\|x\right\|^{2}$ and 
$\left\|x^{2}\right\| \leq \left\|x^{2} + y^{2}\right\|$
for any $x,y \in A$. The subset $A_+ := \left\{x^{2} \: | \: x \in A\right\} $ 
of a JB algebra $A$ is a closed convex cone, and
a partial ordering is defined via $x \leq y :\Leftrightarrow y - x \in A_+$.
Then $\left\{y,x,y\right\} \geq 0$ for any $x \in A_+$ and any $y \in A$.

A \textit{JBW-algebra} is a JB-algebra that is the dual of a Banach space.
Any JBW-algebra has a unit denoted by $\mathbb{I}$. In the finite-dimensional 
case, the JBW-algebras are identical with the JB-algebras 
and with the formally real Jordan algebras.

Finite-dimensional formally real Jordan algebras are the matrix algebras 
$H_n(\mathbb{R})$, $H_n(\mathbb{C})$, $H_n(\mathbb{H})$ ($n = 2,3,4,...$) 
and $H_3(\mathbb{O})$. They consist of the self-adjoint $n$$\times$$n$-matrices
over the real numbers($\mathbb{R}$), the complex numbers ($\mathbb{C}$),
the \emph{quaternions} ($\mathbb{H}$) and
the \emph{octonions} ($\mathbb{O}$) with the usual Jordan product
$x \circ y := (xy + yx)/2$.
The quaternions and octonions are also called \emph{Hamilton numbers} and
\emph{Cayley numbers}, respectively.
For $x,y$ in $H_n(\mathbb{R})$, $H_n(\mathbb{C})$ or $H_n(\mathbb{H})$,
the Jordan triple product $\left\{x,y,x\right\}$ coincides with the simple 
matrix product $xyx$, since $\mathbb{R}$, $\mathbb{C}$ and $\mathbb{H}$ are associative.
However, this does not hold for $x,y$ in $H_3(\mathbb{O})$,
because the octonions are not associative.
Furthermore, there are the \emph{spin factors} or \emph{type $I_2$-factors};
examples for them are $H_2(\mathbb{R})$, $H_2(\mathbb{C})$, $H_2(\mathbb{H})$ 
and $H_2(\mathbb{O})$, but there are many more (including infinite-dimensional ones).
Every finite-dimensional formally real Jordan algebra
can be decomposed into a direct sum of spin factors and 
matrix algebras of the above types \cite{hanche1984jordan, von1933algebraic}.

With the order relation $\leq$ defined by the cone $A_+$, 
the idempotent elements (projections) in a JBW-algebra $A$
form an orthomodular lattice $L_A$ (projection lattice) and thus
$$L_A := \left\{ p \in A \: | \: p^{2} = p\right\}$$
becomes a quantum logic; its orthocomplementation is
$p' := \mathbb{I} - p$ for $p \in L_A$.
For any $p,q \in L_A$, we have: 
$\left\{p,q,p\right\} = 2 p \circ (p \circ q) - p \circ q$;
$p \leq q$ iff $p \circ q = p$ iff $\left\{p,q,p\right\} = p$ 
iff $\left\{q,p,q\right\} = p$; 
$p$ and $q$ are orthogonal iff $p \circ q = 0$ 
iff $\left\{p,q,p\right\} = 0$ 
iff $\left\{q,p,q\right\} = 0$ \cite{AS02, hanche1984jordan}. 
Moreover, two elements $p,q \in L_A$ are compatible
iff they operator-commute (this means: 
$ p \circ (q \circ x) = q \circ (p \circ x)$ 
for all $x \in A$ \cite{AS02}).

A linear functional $\mu: A \rightarrow \mathbb{R}$ is called positive, 
if $\mu(A_+) \subseteq [0,\infty[$. 
For each $0 \neq x \in A$ there is such a 
positive linear functional $\mu$ with $\mu(x) \neq 0$.
The restrictions of the positive
linear functionals to $L_A$ provide a strong 
(see footnote 2) state space $S_A$ for 
the quantum logic $L_A$. It is this 
natural set of states that will always be used 
in the remaining part of this paper.

If none of the direct summands 
in the decomposition of the JBW-algebra $A$
is a spin factor, $S_A$ includes \emph{all} states on $L_A$; 
this is the Gleason theorem for JBW-algebras \cite{bunce1989continuity}.
There is a spin factor for each cardinality except 0, 1 and 2.
The most simple one is the $H_2(\mathbb{R})$; 
it has the real dimension three. The next one is $H_2(\mathbb{C})$
with the real dimension four; it is the self-adoint part of $M_2(\mathbb{C})$
(the 2$\times$2-matrices over the complex numbers $\mathbb{C}$).
The matrix algebra $M_2(\mathbb{C})$ represents the 
two-dimensional version of usual quantum mechanics 
with the complex numbers and is the model for the
spin 1/2 or for the single qubit in quantum information theory.
Only to include these cases, the effort with the
distinction between the positive linear functionals 
and the set of \emph{all} states on the quantum logic $L_A$ is made here.
The other spin factors are more exotic and less interesting.
The distinction becomes needless 
when the algebras of $n$$\times$$n$-matrices with $n \neq 2$ 
(or Hilbert spaces with dimension $n \neq 2$) are considered.

In the following sections, a little knowledge of the 
\emph{Shirshov-Cohn theorem} and some further results 
from the theory of the JB- and JBW-algebras
will be required. We shall go into 
this at those places 
where it will be needed. 
For more information,
it is referred to Refs. \cite{AS02, hanche1984jordan}.

\section{Transition probability in Jordan algebras I}

The following proposition shows how
the transition probability in the 
quantum logic $L_A$ of a JBW-algebra $A$ is linked to
the algebraic structure of $A$.
\vspace{0,3 cm}

\textbf{Proposition 7.1}
\itshape
Suppose that $p \neq 0$ and $q$ are elements 
in the quantum logic $L_A$ of any JBW-algebra $A$. 
The following statements are equivalent:
\begin{enumerate}
\item[\normalfont(a)] 
\itshape
The transition probability from $p$ to $q$ exists and
$\mathbb{P}(q|p) = s$.
\item[\normalfont(b)] 
\itshape
$p$ and $q$ satisfy 
the algebraic identity $\left\{p,q,p\right\} = s p$.
\end{enumerate}
\normalfont

\textbf{Proof.}
For any state $\mu \in S_A$ and $x,y \in A$, 
the following Cauchy-Schwarz inequality holds
(see e.g. 3.6.2 in \cite{hanche1984jordan}):
$$\left|\mu(x \circ y)\right| \leq \left(\mu(x^{2}\right)^{1/2} \left(\mu(y^{2}\right)^{1/2}.$$
This implies that, for $y \in L_A$ with $\mu(y)=0$,
$\mu(x\circ y) = \mu(y \circ x) = 0$ for all $x \in A$. 
Note that $y^{2} = y$ holds for $y \in L_A$.

(b) $\Rightarrow$ (a):
Suppose $\left\{p,q,p\right\} = s p$. 
Note that $ q = \left\{p,q,p \right\} + 2 p' \circ (p \circ q) + p' \circ q$.
If $\mu \in S_A$ and $\mu(p)=1$, then $\mu(p')=0$ and 
$\mu(q) = \mu(\left\{p,q,p\right\}) = s \mu(p) = s$.
Therefore $\mathbb{P}(q|p) = s$.

(a) $\Rightarrow$ (b):
Now suppose $\mathbb{P}(q|p) = s$ and let $\mu$ be any state in $S_A$.
If $\mu(p) = 0$, then $\mu(\left\{p,q,p\right\}) = 2 \mu(p \circ (p \circ q)) - \mu(p \circ q) = 0 = \mu(sp)$.
If $\mu(p) > 0$, define a state $\mu_p \in S_A$ by 
$$\mu_p(x)=\frac{1}{\mu(p)} \mu(\left\{p,x,p\right\})$$ for $x \in A$. 
Then $\mu_p(p) = 1$ and therefore

$$s = \mu_p(q) =\frac{1}{\mu(p)} \mu(\left\{p,q,p\right\}).$$ We have 
$\mu(\left\{p,q,p\right\}) = s \mu(p)$ for all $\mu \in S_A$
and thus $\left\{p,q,p\right\} = sp$.\footnote{Note that, 
with $s=1$, this also shows that $S_A$ is strong,
since $\left\{p,q,p\right\} = p$ iff $p \leq q$.}
\hfill $\square$
\vspace{0,3 cm}

We now come back to Lemma 3.2 (iv) and show 
that its reverse implication is true in the JBW-algebras.
\vspace{0,3 cm}

\textbf{Corollary 7.2} 
\itshape
Suppose that $p$ is an atom (minimal element)
in the quantum logic $L_A$ of any JBW-algebra $A$.
Then $\mathbb{P}(q|p)$ exists for all $q \in L_A$.
\normalfont
\vspace{0,3 cm}

\textbf{Proof.} This follows from Proposition 7.1. Note that
$\left\{\left\{p,x,p\right\}|x \in A \right\}$ $= \mathbb{R} p$ holds for the atoms $p$
in the projection lattice of a JBW-algebra (see e.g. 3.29 in \cite{AS02}).
\hfill $\square$
\vspace{0,3 cm}

The self-adjoint part of a von Neumann algebra 
on a Hilbert space $H$ with inner product $\left\langle \; | \; \right\rangle$
becomes a JBW-algebra $A$ with $x \circ y := (xy+yx)/2$
for $x,y \in A$. 
Here the Jordan triple product $\left\{y,x,y\right\}$
coincides with the operator product $yxy$.
If $\xi$ and $\psi$ are two normalized elements in $H$
and if the projections $p$ and $q$ on the 
one-dimensional linear subspaces that 
$\xi$ and $\psi$ each generate belong to $A$, we get
\begin{center}
$pqp = \left|\left\langle  \xi | \psi \right\rangle\right|^{2} p$
and 
$qpq = \left|\left\langle  \xi | \psi \right\rangle\right|^{2} q$
\end{center}
and therefore
\begin{center}
$\mathbb{P}(q|p) = \mathbb{P}(p|q)= \left|\left\langle  \xi | \psi \right\rangle\right|^{2}$.
\end{center}
This discloses the relation between Definition 3.1
and the usual quantum mechanical transition probability.
If $p$ remains as above, but $q$ is any projection in $A$, we get
$$pqp = \left\langle \xi |q \xi  \right\rangle p$$ and 
$$\mathbb{P}(q|p) = \left\langle \xi |q \xi  \right\rangle.$$
In this way, $p$ then defines the pure state
$q \rightarrow \mathbb{P}(q|p) = \left\langle \xi |q \xi  \right\rangle$.
However, the existence of $\mathbb{P}(q|p)$ for some, 
but not for all projections $q$ in $A$ 
does not require that $p$ is a projection on 
a one-dimensional subspace or an atom. 
An explicit example with real 4$\times$4-matrices
was already presented in Ref. \cite{nie2020alg_origin}.
Some more general examples shall now be introduced.
All these examples demonstrate how
the transition probability of Definition 3.1
differs from the usual quantum mechanical version
and the approaches in Refs.
\cite{AS02,guz1980non, Landsman1997, maczynski1981, mielnik1969theory, mielnik1974generalized, 
pulmannova1989ql_and_trans_prob, vNgeotp}.

Consider a $m$$\times$$n$-matrix $u$ ($m,n \in \mathbb{N}$) with 
entries from $\mathbb{R}$, $\mathbb{C}$ or $\mathbb{H}$
and $uu^{*} = \mathbb{I}_m$.
Here $u^{*}$ denotes the transpose of $u$ in the real case
and the conjugate transpose of $u$ in the other cases; 
$\mathbb{I}_m$ is the identity matrix of size $m \times m$.
Then $u^{*}u$ is an $n$$\times$$n$-matrix with
$(u^{*}u)^{2} = u^{*}(uu^{*})u = u^{*}u$; this requires $m \leq n$.
Now choose any real number $0 \leq s \leq 1$ and define the following two matrices
\begin{center}
$
p := \left(
\begin{array}{cccc}
 \mathbb{I}_m & 0  \\
   &    \\
 0 & 0  \\
\end{array}
\right)
$
and
$
q := \left(
\begin{array}{cccc}
 s \mathbb{I}_m            & s^{1/2}(1-s)^{1/2} u \\
                           &                      \\
 s^{1/2}(1-s)^{1/2} u^{*}  & (1-s) u^{*}u         \\
\end{array}
\right)
$
\end{center}
in $H_{m+n}(K)$, $K \in \left\{ \mathbb{R}, \mathbb{C}, \mathbb{H} \right\}$.
Some simple matrix calculations yield $p^{2} = p$, $q^{2} = q$, $pqp = s p$
and $qpq = s q$. By Proposition 7.1 $\mathbb{P}(q|p)$ and $\mathbb{P}(p|q)$
both exist and we have $\mathbb{P}(q|p) = \mathbb{P}(p|q) = s$.
Selecting $0 \neq s \neq 1$ and $m \geq 2$ provides many examples of 
non-trivial transition probabilities $\mathbb{P}(q|p)$
where $p$ is not an atom. Using any unitary element
$w \in H_{m+n}(K)$ (this means $w^{*}w = ww^{*} = \mathbb{I}_{m+n}$), further examples 
where $p$ gets a different form
can be constructed by replacing $p$ and $q$
with $w^{*}pw$ and $w^{*}qw$.

\section{Isoclinic pairs in Jordan algebras}

In the above example, the projections $p$ and $q$
satisfy the two identities $pqp = s p$ and $qpq = s q$ with $0 \leq s \leq 1$.
In a von Neumann algebra $M$, such a pair 
is called \textit{isoclinic} \cite{christensen1982measures, hamhalter2015dye, maeda1989probability}.
This algebraic property is equivalent to a geometric property called 
\textit{equiangularity} \cite{Freedman_2019}.
The subalgebra of $M$ that an isoclinic pair generates 
is an isomorphic copy of the matrix algebra $M_2(\mathbb{C})$
(the 2$\times$2-matrices over the complex numbers $\mathbb{C}$).
By Proposition 7.1, the pair $p$ and $q$
is isoclinic iff the transition probabilities
$\mathbb{P}(q|p)$ and $\mathbb{P}(p|q)$ both exist and coincide. 
We are now going to study this situation in the Jordan algebraic framework.

With any elements $a_1, ..., a_n$ in a Jordan algebra $A$, 
$A_{\left\{a_1, ..., a_n\right\}}$ shall denote
the Jordan subalgebra generated by $a_1, ..., a_n$. 
Note that this subalgebra need not
include the unit element $\mathbb{I}$ of $A$.

By the Shirshov-Cohn theorem \cite{hanche1984jordan}, 
any Jordan algebra generated by the unit element $\mathbb{I}$
and two further elements $a_1$ and $a_2$ 
is \emph{special}. This means that we can assume 
that $A_{\left\{\mathbb{I},a_1,a_2\right\}}$ is a part of an associative algebra 
and that the product in $A_{\left\{\mathbb{I},a_1,a_2\right\}}$
can be derived from this associative algebra 
in the following way: 
$x \circ y = \frac{1}{2}(xy + yx)$ for any $x,y \in A_{\left\{\mathbb{I},a_1,a_2\right\}}$. 
Then $\left\{ x,y,x \right\} = xyx$. This holds as well for 
$A_{\left\{a_1,a_2\right\}} \subseteq A_{\left\{\mathbb{I},a_1,a_2\right\}}$
and will be very useful
for the study of the Jordan algebra generated by two projections
with existing transition probability.

Note that an element $v$ in a JBW-algebra $A$ 
is called a \textit{symmetry} if $ v^{2} = \mathbb{I}$
\cite{hanche1984jordan} and that $H_2(\mathbb{R})$
denotes the Jordan algebra that consists of the 
self-adjoint 2$\times$2-matrices over the real numbers. 
\vspace{0,3 cm}

\textbf{Lemma 8.1} 
\itshape
Suppose that the transition probabilities $\mathbb{P}(q | p)$ 
and $\mathbb{P}(p | q)$ both exist for two non-zero elements $p$ and $q$ 
in the quantum logic $L_A$ of any JBW-algebra $A$. 
\begin{enumerate}
\item[\normalfont(i)] 
Then $\mathbb{P}(q | p) = \mathbb{P}(p | q)$.
\item[\normalfont(ii)]
If $\mathbb{P}(q | p) \neq 0 \neq \mathbb{P}(p | q)$,
there is a symmetry $v \in A_{\left\{\mathbb{I},a_1,a_2\right\}} \subseteq A$
such that $q = \left\{v,p,v\right\}$ and $p = \left\{v,q,v\right\}$.
\item[\normalfont(iii)] 
$\mathbb{P}(q | p) = \mathbb{P}(p | q) = 1$ iff $p = q$.
In this case, $A_{\left\{p,q\right\}} = \mathbb{R} p$.
\item[\normalfont(iv)] 
$\mathbb{P}(q | p) = \mathbb{P}(p | q) = 0$ 
iff $p$ and $q$ are orthogonal.
In this case, 
$A_{\left\{p,q\right\}} = \mathbb{R} p \oplus \mathbb{R} q$.
\item[\normalfont(v)] 
$ 0 < \mathbb{P}(q | p) = \mathbb{P}(p | q) < 1 $ iff
$p$, $q$ and $p \circ q$ are linearly independent.
The three elements $p$, $q$ and $p \circ q$ 
generate $A_{\left\{p,q\right\}}$ and, in this case, 
$A_{\left\{p,q\right\}}$ is isomorphic to $H_2(\mathbb{R})$. 
\end{enumerate}
\normalfont

\textbf{Proof.}
Consider the subalgebras 
$A_{\left\{p,q\right\}} \subseteq A_{\left\{\mathbb{I},p,q\right\}}$ 
of $A$. By the Shirshov-Cohn theorem, we can assume 
that the Jordan product on them
stems from an associative product of some larger algebra.

(i) Define $r := \mathbb{P}(q | p)$ and $ s = \mathbb{P}(p | q)$.
Then $pqp = rp$ and $qpq = sq$ by Proposition 7.1 and we get
$rpq = pqpq = s pq$. If $pq \neq 0$, then $r = s$. 
If $pq = 0$, then $pqp = 0$ and $qpq = 0$; therefore $r = s = 0$.
Thus we have (i).

(ii) will be proved later, since (v) will be needed. Items
(iii) and (iv) follow immediately from the general properties 
of the transition probabilities (see section 3).

(v) If $p$, $q$ and $p \circ q$ are linearly independent,
the cases $p \circ q = 0$ and $p = q$ are ruled out.
The first one is identical with the case that 
$p$ and $q$ are orthogonal and $\mathbb{P}(q | p) = 0$.
The second one is identical with the case that 
$p \leq q$ and $ q \leq p$, which is the same as
$\mathbb{P}(q | p) = 1$ and $\mathbb{P}(p | q) = 1$.

Now suppose $\mathbb{P}(q | p) = \mathbb{P}(p | q) = s$
and $0 \neq s \neq 1$;
by Proposition 7.1 
this means $pqp = s p$ and $qpq = s q$. 

Using $A_{\left\{\mathbb{I},p,q\right\}}$ and the 
Shirshov-Cohn theorem, we first show the 
linear independence of $p$, $q$ and $p \circ q$.
Suppose $0 = r_1 p + r_2 q + r_3 p \circ q$ 
with $r_1, r_2, r_3 \in \mathbb{R}$. Then
$0 = q'(r_1 p + r_2 q + r_3 p \circ q)q' = r_1 q'pq'$.
We have $q'pq' \neq 0$, since otherwise $p$ and $q'$ 
are orthogonal and thus $p \leq q$ and $s = 1$.
Therefore $r_1 = 0$. 
Using $p'$ instead of $q'$, it follows in the same way 
that $r_2 = 0$. Finally $r_3 = 0$, since 
$p \circ q = 0$ would mean that 
$p$ and $q$ become orthogonal and then $s = 0$.

Furthermore, from $sp = pqp = 2 p \circ (p \circ q) - p \circ q$ we get 
$$p \circ (p \circ q) = (sp + p\circ q)/2,$$
and from 
$sq = qpq = 2 q \circ (p \circ q) - p \circ q$ we get 
$$q \circ (p \circ q) = (sq + p \circ q)/2.$$
Moreover 
\begin{align*}
\left(p \circ q\right)^{2} & = \left( pq + qp \right)^{2}/4 
   = \left( pqpq + pqp + qpq + qpqp\right)/4 \\
 & = \left( spq + sp + sq + sqp \right)/4 \\
 & = \left( 2s p\circ q + sp + sq \right)/4 .
\end{align*}
Therefore $p \circ (p \circ q)$, $q \circ (p \circ q)$ 
and $(p \circ q)^{2}$ lie in the linear hull of 
$p$, $q$ and $p \circ q$ which thus becomes a three-dimensional
Jordan algebra and identical with $A_{\left\{p,q\right\}}$.
Now consider the following matrices in $H_2(\mathbb{R})$:
\begin{center}
$a := 
\begin{pmatrix}
1 & 0 \\
  &   \\
0 & 0
\end{pmatrix}$
and
$b := 
\begin{pmatrix}
s                  & s^{1/2}(1-s)^{1/2} \\
                   &                    \\
s^{1/2}(1-s)^{1/2} & 1-s
\end{pmatrix}$.
\end{center}
Then $a^{2} = a$, $b^{2} = b$, $aba = sa$ and $bab = sb$.
Note that these matrices are rather simple versions of those
used in the example at the end of section 7 with
$K = \mathbb{R}$, $m = n = 1$ and $u = 1$.
The same line of reasoning as above with $p$ and $q$ 
or some simple matrix calculations then show that 
$a$, $b$ and $a \circ b$ 
become linearly independent 
with $0 \neq s \neq 1$ and that
the following identities hold:
\begin{align*}
a \circ (a \circ b) & = (sa + a\circ b)/2, \\
b \circ (a \circ b) & = (sb + a\circ b)/2 \; \text{and} \\
\left(a \circ b\right)^{2} & = (2s a\circ b + sa + sb)/4.
\end{align*}

Now define a linear map 
$\pi : A_{\left\{p,q\right\}} \rightarrow H_2(\mathbb{R})$
by $\pi (r_1 p + r_2 q + r_3 p \circ q ) := r_1 a + r_2 b + r_3 a \circ b$
for $r_1, r_2, r_3 \in \mathbb{R}$.
Since $p$, $q$ and $p \circ q$ are linearly independent, 
$\pi$ is well-defined.
The above identities 
for $p$ and $q$ and for $a$ and $b$
imply that $\pi$ is multiplicative.
Since the real
dimension of $H_2(\mathbb{R})$ is three, the Jordan algebras
$A_{\left\{p,q\right\}}$ and $H_2(\mathbb{R})$ 
are isomorphic. 

(ii) By Proposition 7.1 and part (i) of this lemma, 
we can assume that $pqp = sp$ and $qpq = sq$ with 
$s = \mathbb{P}(q | p) =\mathbb{P}(p | q) \neq 0$.
With part (v) of this lemma, we can conclude that 
$A_{\left\{\mathbb{I},p,q\right\}}$ has a finite dimension;
it thus becomes a JBW-algebra.  
By Lemma 5.2.1 in Ref. \cite{hanche1984jordan},
there is a symmetry $v \in A_{\left\{\mathbb{I},p,q\right\}}$
such that $vpqpv = qpq$. Then
$sq = qpq = vpqpv = s vpv $ and thus $q = vpv$.
Furthermore, $vqv = v^{2}pv^{2} = p$.
\hfill $\square$
\vspace{0,3 cm}

When we apply Lemma 8.1 (ii) to the usual quantum mechanical setting
or to the self-adjoint part of a von Neumann algebra, it tells us that the
two projections $p$ and $q$ must be 
unitarily equivalent, if $\mathbb{P}(q | p)$ and 
$\mathbb{P}(p | q)$ both exist and 
$\mathbb{P}(q | p) \neq 0 \neq \mathbb{P}(p | q)$ holds.

The complex $\ast$-algebra that
such an isoclinic projection pair $p$ and $q$ in a von Neumann algebra
generates is a copy of $M_2(\mathbb{C})$, but the Jordan
algebra $A_{\left\{p,q\right\}} \cong H_2(\mathbb{R})$ 
that it generates, by Lemma 8.1 (v), is smaller than the
self-adjoint part of $M_2(\mathbb{C})$.
The self-adjoint part of $M_2(\mathbb{C})$ 
has the real dimension 4, while $H_2(\mathbb{R})$
has the real dimension 3.
Therefore, the Jordan algebra that an 
isoclinic pair in a von Neumann algebra
generates provides more structural information
about the pair than than complex $\ast$-algebra that
it generates.

\section{Transition probability in Jordan algebras II}

We now return to the case when only $\mathbb{P}(q | p)$ exists 
for the elements $p \neq 0$ and $q$ in the quantum logic of a JBW-algebra.
The following proposition shows how this is still related to the isoclinicity.
\vspace{0,3 cm}

\textbf{Proposition 9.1} 
\itshape
Suppose that the transition probability $\mathbb{P}(q | p)$ exists 
for two elements $p \neq 0$ and $q$ in the quantum logic $L_A$ of any 
JBW-algebra $A$ and that $\mathbb{P}(q | p) \neq 0$. Then there 
are two orthogonal elements $q_o$ and $q_1$ in $L_A$ such that
\begin{enumerate}
\item[\normalfont(a)] $q = q_o + q_1$,
\item[\normalfont(b)] $q_1$ and $p$ are orthogonal, and
\item[\normalfont(c)] $\mathbb{P}(q_o | p)$ as well as $\mathbb{P}(p | q_o)$ 
           exist and both coincide with $\mathbb{P}(q | p)$.
           This means that $p$ and $q_o$ form an isoclinic pair.
\end{enumerate}
\normalfont

\textbf{Proof.} 
Use the Shirshov-Cohn theorem again and
consider the subalgebra $A_{\left\{p,q\right\}}$
embedded in a larger associative algebra 
defining the special Jordan product 
on $A_{\left\{p,q\right\}}$.
Define $s := \mathbb{P}(q | p)$ and 
$q_o := \frac{1}{s} qpq $.
By Proposition 7.1 we have 
the identity $pqp = sp$, which
will be used repeatedly in the following equations:
$${q_o}^{2} = \frac{1}{s^{2}} qpqpq = \frac{1}{s} qpq = q_o$$ 
and therefore $q_o \in L_A$.
$$p q_o p = \frac{1}{s} pqpqp = \frac{1}{s} (pqp)^{2} = sp$$
and therefore $\mathbb{P}(q_o | p) = s$ by Proposition 7.1. 
$$q_o p q_o = \frac{1}{s^{2}} qpqpqpq = \frac{1}{s^{2}} q(pqp)(pqp)q = qpq = s q_o$$
and therefore $\mathbb{P}(p | q_o) = s$ by Proposition 7.1.
Moreover $$q_o \circ q = \frac{1}{2s} (qpq^{2} + q^{2}pq) = \frac{1}{2s} (qpq + qpq) = q_o$$
and therefore $q_o \leq q$.
Now define $q_1 := q - q_o \in L_A$. Then 
$$q_1 \circ p = q\circ p -q_o \circ p = q\circ p - \frac{1}{2s}(qpqp + pqpq) 
= q\circ p - \frac{1}{2}(qp + pq) = 0$$
and therefore $q_1$ and $p$ are orthogonal.
\hfill $\square$
\vspace{0,3 cm}

A \textit{trace} on a JBW-algebra $A$ is a map 
$ \tau : A_+ \rightarrow [0,\infty]$ with
$ \tau \left(r x + s y\right) =  r \: \tau \left(x\right) + s \: \tau \left(y\right)$
for $x,y \in A_+$, $r,s \in \mathbb{R}$, $0 \leq r,s$, 
and $\tau(\left\{x,y^{2},x\right\}) = \tau(\left\{y,x^{2},y\right\})$ 
for all $x,y \in A$ \cite{bunce1995traces}. 
For the projections $p,q \in L_A$ we then have
$ \tau (\left\{p,q,p\right\}) = \tau (\left\{q,p,q\right\}) $.
\vspace{0,3 cm}

\textbf{Corollary 9.2}
\itshape
Let $A$ be a JBW-algebra, $\tau$ a trace on $A$ and $p,q \in L_A$, $0 \neq p,q$.
\begin{enumerate}
\item[\normalfont(i)]
If $\mathbb{P}(q|p)$ and $\mathbb{P}(p|q)$ both exist 
and $\mathbb{P}(q|p) = \mathbb{P}(p|q) \neq 0$, 
then $\tau(p) = \tau(q)$.
\item[\normalfont(ii)]
If $\mathbb{P}(q|p)$ exists and $\mathbb{P}(q|p) \neq 0$, 
then $\tau(p) \leq \tau(q)$.
\end{enumerate}
\normalfont

\textbf{Proof.}
(i) $\mathbb{P}(q|p) = \mathbb{P}(p|q) = s \neq 0$, i.e. 
$\left\{p,q,p\right\} = sp$ and $\left\{q,p,q\right\} = sq$
by Proposition 7.1. Then 
$s \: \tau(p) = \tau(\left\{p,q,p\right\}) =\tau(\left\{q,p,q\right\}) = s \: \tau(q)$
and thus $ \tau(p) = \tau(q) $. Another way to prove this is to use Lemma 8.1 (ii).

(ii) With $q_o$ and $q_1$ as in Proposition 9.1,
wet get from (i): 
$\tau(p) = \tau(q_o) \leq \tau(q_o +q_1) = \tau(q)$.
\hfill $\square$
\vspace{0,3 cm}

As an example, consider the JBW-algebra, consisting of the
self-adjoint linear operators on a Hilbert space $H$,
with the usual trace. In this case,
the trace of a projection $p$ is identical with the dimension 
of the associated closed linear subspace $pH$. 
Corollary 9.2 then tells us that
$dim(pH) = dim(qH)$ must hold
if $\mathbb{P}(q|p)$ and $\mathbb{P}(p|q)$ both exist 
and $\mathbb{P}(q|p) = \mathbb{P}(p|q) \neq 0$, 
and that $dim(pH) \leq dim(qH)$ must hold 
if $\mathbb{P}(q|p)$ exists and $\mathbb{P}(q|p) \neq 0$. 
This can also be derived 
directly from Proposition~7.1 in a simpler way: 
$pqp = sp$ with $s \neq 0$ implies $ pH = pqpH \subseteq pqH$
and therefore $dim(pH) \leq dim(pqH) \leq dim(qH)$.
In the general Jordan algebraic setting, however,
this line of reasoning will not work.

The following theorem finally provides a complete characterization
of the Jordan algebra $A_{\left\{p,q\right\}}$ 
generated by a pair of projections $p \neq 0$ and $q$ 
for that $\mathbb{P}(q | p)$ exists.
\vspace{0,3 cm}

\textbf{Theorem 9.3}
\itshape
Suppose that the transition probability $\mathbb{P}(q | p)$ exists 
for two elements $p \neq 0$ and $q$ in the quantum logic $L_A$ of any 
JBW-algebra $A$.
\begin{enumerate}
\item[\normalfont(i)] If $\mathbb{P}(q | p) = 0$, 
then $A_{\left\{p,q\right\}} = \mathbb{R} p \oplus \mathbb{R} q$.
\item[\normalfont(ii)] If $\mathbb{P}(q | p) = 1$, 
then $A_{\left\{p,q\right\}} = \mathbb{R} p \oplus \mathbb{R} (q-p)$.
\item[\normalfont(iii)] If $0 < \mathbb{P}(q | p) < 1$, 
then $A_{\left\{p,q\right\}} = A_{\left\{p,q_o\right\}} \oplus \mathbb{R} (q - q_o)$,
where $q_o$ is an element in $L_A$ with $0 \neq q_o \leq q$ such that
$\mathbb{P}(q_o | p)$ as well as $\mathbb{P}(p | q_o)$ 
exist and both coincide with $\mathbb{P}(q | p)$;
$p$ and $q - q_o$ are orthogonal, and
$A_{\left\{p,q_o\right\}}$ is isomorphic to $H_2(\mathbb{R})$.
\end{enumerate}
\normalfont

\textbf{Proof.} Part (i) follows from the orthogonality of $p$ and $q$ 
in the case when $\mathbb{P}(q | p) = 0$.
Part (ii) follows from the inequality $p \leq q$ 
in the case when $\mathbb{P}(q | p) = 1$.
Part (iii) follows by combining Lemma 8.1 (v) and Proposition 9.1.
\hfill $\square$

\section{Conclusions}

An extension of the usual quantum mechanical transition probability
to a very general setting has been presented. These are the
quantum logics the mathematical structure of which is an
orthomodular partially ordered set.

An interesting aspect of the transition probability
considered here
is that it does not require any state,
but it has a purely algebraic origin. The transition 
probability $\mathbb{P}(q|p)$, if it exists,
becomes a characteristic of the algebraic relation 
between two elements $p \neq 0$ and $q$ in the quantum logic;
$p$ needs not be an atom and no state can then be allocated to $p$.
This is a major difference from other approaches
\cite{AS02,guz1980non, Landsman1997, maczynski1981, 
mielnik1969theory, mielnik1974generalized, 
pulmannova1989ql_and_trans_prob, vNgeotp}.

If $p$ and $q$ are compatible, only three cases are possible 
for the transition probability: either
it does not exist or $\mathbb{P}(q|p) =1$, 
which is equivalent to $p \leq q$, 
or $\mathbb{P}(q|p) =0$, which is equivalent to
the orthogonality $p$ and $q$. 
Only the same three cases would be possible, if $p$ and $q$ were
elements in a Boolean algebra. The inequality $p \leq q$ 
is a logical relation between the propositions 
$p$ and $q$; it means that $p$ implies $q$. 
The orthogonality of of $p$ and $q$ is another logical relation 
between the propositions $p$ and $q$; it means that $p$ rules out $q$. 
Therefore, $\mathbb{P}(q|p)$ 
can be considered an extension of these two logical relations 
to certain pairs $p$ and $q$ that are not compatible.
This extended relation, however, is associated with a probability and
introduces a continuum of new cases ($0 < \mathbb{P}(q|p) < 1$) between 
the two classical cases `$p$ implies $q$' ($\mathbb{P}(q|p) = 1$) 
and `$p$ rules out $q$' ($\mathbb{P}(q|p) = 0$).

The no-cloning theorem 
\cite{dieks1982communication, wootters1982single}
plays an
important role in quantum information theory and 
particularly in quantum cryptography.
Theorem~5.1 becomes a very general version of this
theorem in the quantum logical setting and
still covers new cases in common Hilbert space quantum mechanics,
neither requiring the tensor product 
nor the atomic elements in the quantum logic.
Particularly the non-atomic case
goes beyond the usually considered situation and 
creates promising new opportunities for the quantum
key distribution protocols~\cite{BB84, E91}.

The approach presented here 
has enabled us to see that 
the transition probabilities, their invariance under morphisms
and the product rule are sufficient to derive the no-cloning theorem
in the quantum logical framework without falling back to 
Hilbert spaces.
This approach might be considered
rather general and abstract,
but such approaches may be needed
to identify and understand the deeper origins
of the quantum mysteries. 
Searching for their origins
in the common quantum mechanical Hilbert space formalism
can be as difficult as finding 
a needle in the haystack.

Another rewarding approach with a different focus are 
the generalized probabilistic theories used in Refs.
\cite{barnum2006cloning, barnum2007generalized}.
Instead of the quantum logic and its algebraic structure, 
their starting point are the state space and its convex structure,
but it is not evident how the transition probability 
can be defined in that framework. 
One possibility is to construct a quantum logic from the 
\emph{projective units} of Alfsen and Shultz's theory
\cite{alfsen1976non, AS02} first and to use the same definition 
as here then.

An interesting class of quantum logics are
the projection lattices of the JBW-algebras.
We have seen that they
provide the appropriate framework for a structural
analysis (sections 8 and 9) of the transition 
probability and how it
is linked to the Jordan
algebraic structure (Proposition~7.1).
Beyond the usual quantum mechanical model based on the complex
Hilbert space or von Neumann algebras, the JBW-algebras
include versions based on the real numbers or the quaternions
\cite{AS02, hanche1984jordan},
and the no-cloning theorem remains valid in these cases,
although a reasonable tensor product is not 
available \cite{nie2020loc_tomography, wootters1986quantum, wootters1990local}.

Furthermore, there is the
exceptional Jordan algebra $H_3(\mathbb{O})$ 
\cite{albert1934, AS02, baez2002octonions, hanche1984jordan, von1933algebraic},
which is not special (see section 8),
since the product of the octonions is not associative,
and which cannot be represented as linear operators
on any kind of Hilbert space. Nevertheless,
its quantum logic possesses the transition probabilities.
It contains many atoms and many non-orthogonal pairs of atoms; 
by Proposition 7.2, the transition probability then exists in many cases
and non-orthogonal pairs of non-identical atoms result in 
non-trivial transition probabilities.
Explicit examples can be constructed using the same matrices $p$ and $q$
as at the end of section 7, but now with entries from $\mathbb{O}$,
$m = 1$ and $n = 2$. Note that any two octonions (here the components of $u$) 
and their conjugates generate an associative subalgebra of $\mathbb{O}$
\cite{baez2002octonions}. The automorphisms of $H_3(\mathbb{O})$ 
map the matrices $p$ and $q$ to further examples.

\bibliographystyle{abbrv}
\bibliography{Literatur}

\begin{thebibliography}{10}

\bibitem{albert1934}
A.~A. Albert.
\newblock On a certain algebra of quantum mechanics.
\newblock {\em Annals of Mathematics}, pages 65--73, 1934.

\bibitem{alfsen1976non}
E.~M. Alfsen and F.~W. Shultz.
\newblock Non-commutative spectral theory for affine function spaces on convex
  sets.
\newblock {\em Memoirs of the American Mathematical Society}, 172, 1976.

\bibitem{AS02}
E.~M. Alfsen and F.~W. Shultz.
\newblock {\em Geometry of state spaces of operator algebras}.
\newblock Birkh\"auser, Basel, Switzerland, 2003.

\bibitem{baez2002octonions}
J.~Baez.
\newblock The octonions.
\newblock {\em Bulletin of the American Mathematical Society}, 39(2):145--205,
  2002.

\bibitem{barnum2006cloning}
H.~Barnum, J.~Barrett, M.~Leifer, and A.~Wilce.
\newblock Cloning and broadcasting in generic probabilistic theories.
\newblock {\em arXiv:quant-ph/0611295}, 2006.

\bibitem{barnum2007generalized}
H.~Barnum, J.~Barrett, M.~Leifer, and A.~Wilce.
\newblock Generalized no-broadcasting theorem.
\newblock {\em Physical Review Letters}, 99(24):240501, 2007.

\bibitem{barnum1996noncommuting}
H.~Barnum, C.~M. Caves, C.~A. Fuchs, R.~Jozsa, and B.~Schumacher.
\newblock Noncommuting mixed states cannot be broadcast.
\newblock {\em Physical Review Letters}, 76(15):2818, 1996.

\bibitem{beltrametti1984logic}
E.~G. Beltrametti and G.~Cassinelli.
\newblock {\em The logic of quantum mechanics}.
\newblock Cambridge University Press, Cambridge, UK, 1984.

\bibitem{BB84}
C.~H. Bennett and G.~Brassard.
\newblock Quantum cryptography: {P}ublic key distribution and coin tossing.
\newblock In {\em Proceedings of IEEE International Conference on Computers,
  Systems and Signal Processing (Bangalore, India, Dec. 1984)}, volume 175,
  page~8, 1984.

\bibitem{birkhoff-vN36}
G.~Birkhoff and J.~von Neumann.
\newblock The logic of quantum mechanics.
\newblock {\em Annals of Mathematics}, 37:823--843, 1936.

\bibitem{born1926}
M.~Born.
\newblock Quantenmechanik der {S}to{\ss}vorg{\"a}nge.
\newblock {\em Zeitschrift f{\"u}r Physik}, 38(11-12):803--827, 1926.

\bibitem{brabec1979compatibility}
J.~Brabec.
\newblock Compatibility in orthomodular posets.
\newblock {\em {\v{C}}asopis pro p{\v{e}}stov{\'a}n{\'\i} matematiky},
  104(2):149--153, 1979.

\bibitem{brabec-ptak1982compatibility}
J.~Brabec and P.~Pt{\'a}k.
\newblock On compatibility in quantum logics.
\newblock {\em Foundations of Physics}, 12(2):207--212, 1982.

\bibitem{PhysRevA.57.2368}
D.~Bru\ss, D.~P. DiVincenzo, A.~Ekert, C.~A. Fuchs, C.~Macchiavello, and J.~A.
  Smolin.
\newblock Optimal universal and state-dependent quantum cloning.
\newblock {\em Physical Review A}, 57:2368--2378, 1998.

\bibitem{bunce1995traces}
L.~Bunce and J.~Hamhalter.
\newblock {Traces and subadditive measures on projections in JBW-algebras and
  von Neumann algebras}.
\newblock {\em Proceedings of the American Mathematical Society},
  123(1):157--160, 1995.

\bibitem{bunce1989continuity}
L.~Bunce and J.~M. Wright.
\newblock Continuity and linear extensions of quantum measures on {J}ordan
  operator algebras.
\newblock {\em Mathematica Scandinavica}, 64:300--306, 1989.

\bibitem{buvzek1996quantum}
V.~Bu{\v{z}}ek and M.~Hillery.
\newblock Quantum copying: {Beyond} the no-cloning theorem.
\newblock {\em Physical Review A}, 54(3):1844, 1996.

\bibitem{christensen1982measures}
E.~Christensen.
\newblock Measures on projections and physical states.
\newblock {\em Communications in Mathematical Physics}, 86(4):529--538, 1982.

\bibitem{clifton2003characterizing}
R.~Clifton, J.~Bub, and H.~Halvorson.
\newblock Characterizing quantum theory in terms of information-theoretic
  constraints.
\newblock {\em Foundations of Physics}, 33(11):1561--1591, 2003.

\bibitem{dieks1982communication}
D.~Dieks.
\newblock Communication by {EPR} devices.
\newblock {\em Physics Letters A}, 92(6):271--272, 1982.

\bibitem{dorninger2014}
D.~Dorninger and H.~L{\"a}nger.
\newblock A note on {B}oolean subsets of orthomodular posets.
\newblock {\em Italian Journal of Pure and Applied Mathematics}, 32:277--282,
  2014.

\bibitem{E91}
A.~K. Ekert.
\newblock Quantum cryptography based on {B}ell's theorem.
\newblock {\em Physical Review Letters}, 67:661--663, 1991.

\bibitem{Freedman_2019}
M.~Freedman, M.~Shokrian-Zini, and Z.~Wang.
\newblock Quantum computing with octonions.
\newblock {\em Peking Mathematical Journal}, 2(3-4):239–273, 2019.

\bibitem{gudder1979stochastic}
S.~P. Gudder.
\newblock {\em Stochastic methods in quantum mechanics}.
\newblock North-Holland, New York, NY, 1979.

\bibitem{guz1980non}
W.~Guz.
\newblock A non-symmetric transition probability in quantum mechanics.
\newblock {\em Reports on Mathematical Physics}, 17(3):385--400, 1980.

\bibitem{haag1964algebraic}
R.~Haag and D.~Kastler.
\newblock An algebraic approach to quantum field theory.
\newblock {\em Journal of Mathematical Physics}, 5(7):848--861, 1964.

\bibitem{hamhalter1997statistical}
J.~Hamhalter.
\newblock Statistical independence of operator algebras.
\newblock {\em Annales de l' {Institut Henri Poincar\'e}, physique
  th{\'e}orique}, 67(4):447--462, 1997.

\bibitem{hamhalter2015dye}
J.~Hamhalter.
\newblock Dye's theorem and {G}leason's theorem for {AW*-algebras}.
\newblock {\em Journal of Mathematical Analysis and Applications},
  422(2):1103--1115, 2015.

\bibitem{hanche1984jordan}
H.~Hanche-Olsen and E.~St{\o}rmer.
\newblock {\em Jordan operator algebras}.
\newblock Pitman, London, UK, 1984.

\bibitem{von1933algebraic}
P.~Jordan, J.~von Neumann, and E.~Wigner.
\newblock On an algebraic generalization of the quantum mechanical formalism.
\newblock {\em Annals of Mathematics}, 35:29--64, 1934.

\bibitem{kalmbachorthomodular}
G.~Kalmbach.
\newblock {\em Orthomodular lattices}.
\newblock Academic Press, London, UK, 1983.

\bibitem{Kitajima2015}
Y.~Kitajima.
\newblock Imperfect cloning operations in algebraic quantum theory.
\newblock {\em Foundations of Physics}, 45(1):62--74, 2015.

\bibitem{Landsman1997}
N.~P. Landsman.
\newblock Poisson spaces with a transition probability.
\newblock {\em Reviews in Mathematical Physics}, 09(01):29–57, 1997.

\bibitem{maczynski1981}
M.~Maczy{\'n}ski.
\newblock Commutativity and generalized transition probability in quantum
  logic.
\newblock In {\em Current Issues in Quantum Logic}, pages 355--364. Springer,
  1981.

\bibitem{maeda1989probability}
S.~Maeda.
\newblock Probability measures on projections in von {N}eumann algebras.
\newblock {\em Reviews in Mathematical Physics}, 1(02n03):235--290, 1989.

\bibitem{mielnik1969theory}
B.~Mielnik.
\newblock Theory of filters.
\newblock {\em Communications in Mathematical Physics}, 15(1):1--46, 1969.

\bibitem{mielnik1974generalized}
B.~Mielnik.
\newblock Generalized quantum mechanics.
\newblock {\em Communications in Mathematical Physics}, 37(3):221--256, 1974.

\bibitem{:/content/aip/journal/jmp/50/10/10.1063/1.3245811}
T.~Miyadera and H.~Imai.
\newblock No-cloning theorem on quantum logics.
\newblock {\em Journal of Mathematical Physics}, 50(10):102107, 2009.

\bibitem{Nie1998HPA01}
G.~Niestegge.
\newblock Statistische und deterministische {V}orhersagbarkeit bei der
  quantenphysikalischen {M}essung.
\newblock {\em Helvetica Physica Acta}, 71(2):163--183, 1998.

\bibitem{niestegge2001non}
G.~Niestegge.
\newblock Non-{Boolean} probabilities and quantum measurement.
\newblock {\em Journal of Physics A: Mathematical and General}, 34(30):6031,
  2001.

\bibitem{nie2017QKD}
G.~Niestegge.
\newblock Quantum key distribution without the wavefunction.
\newblock {\em International Journal of Quantum Information}, 15(06):1750048,
  2017.

\bibitem{nie2020loc_tomography}
G.~Niestegge.
\newblock Local tomography and the role of the complex numbers in quantum
  mechanics.
\newblock {\em Proceedings of the Royal Society A}, 476(2238):20200063, 2020.

\bibitem{nie2020alg_origin}
G.~Niestegge.
\newblock Quantum probability's algebraic origin.
\newblock {\em Entropy}, 22(11):1196, 2020.

\bibitem{piron1964axiomatique}
C.~Piron.
\newblock Axiomatique quantique.
\newblock {\em Helvetica Physica Acta}, 37(4-5):439--468, 1964.

\bibitem{ptak1991orthomodular}
P.~Pt{\'a}k and S.~Pulmannov{\'a}.
\newblock {\em Orthomodular structures as quantum logics}.
\newblock Kluwer, Dordrecht, the Netherlands, 1991.

\bibitem{ptak1983measures}
P.~Pt{\'a}k and V.~Rogalewicz.
\newblock Measures on orthomodular partially ordered sets.
\newblock {\em Journal of Pure and Applied Algebra}, 28(1):75--80, 1983.

\bibitem{pulmannova1989ql_and_trans_prob}
S.~Pulmannova.
\newblock Representations of quantum logics and transition probability spaces.
\newblock In {\em The concept of probability}, pages 51--59. Springer,
  Dordrecht, the Netherlands, 1989.

\bibitem{redei1995logical}
M.~R{\'e}dei.
\newblock Logical independence in quantum logic.
\newblock {\em Foundations of Physics}, 25(3):411--422, 1995.

\bibitem{sakai1971}
S.~Sakai.
\newblock {\em C*-algebras and W*-algebras}.
\newblock Springer, Berlin Heidelberg, 1971.

\bibitem{varadarajan1968and1970}
V.~S. Varadarajan.
\newblock {\em Geometry of quantum theory, vols. 1 and 2}.
\newblock Van Nostrand Reinhold, New York, NY, 1968 and 1970.

\bibitem{vNgeotp}
J.~von Neumann.
\newblock Continuous geometries with a transition probability (prepared and
  edited by {Israel Halperin}).
\newblock {\em Memoirs of the American Mathematical Society}, 34(252):210pages,
  1981.

\bibitem{wootters1986quantum}
W.~K. Wootters.
\newblock Quantum mechanics without probability amplitudes.
\newblock {\em Foundations of physics}, 16(4):391--405, 1986.

\bibitem{wootters1990local}
W.~K. Wootters.
\newblock Local accessibility of quantum states.
\newblock In W.~H. Zurek, editor, {\em Complexity, entropy and the physics of
  information}, pages 39--46. Addison-Wesley, Boston, MA, 1990.

\bibitem{wootters1982single}
W.~K. Wootters and W.~H. Zurek.
\newblock A single quantum cannot be cloned.
\newblock {\em Nature}, 299(5886):802--803, 1982.

\end{thebibliography}
\end{document}